# Long-distance Liquid Transport Along Fibers Arising From Plateau-Rayleigh Instability

Yunqiao Huang[a], Xianguo Li[a], Zhongchao Tan[a,b*]

[a] *Department of Mechanical and Mechatronics Engineering, University of Waterloo, 200 University Avenue West, Waterloo, N2L 3G1, Ontario, Canada*

[b] *Eastern Institute of Technology, Ningbo, No.568 Tongxin Road, Zhenhai District, Ningbo, Zhejiang 315200, China*

[*]*Corresponding author: ztan@eitech.edu.cn (Z. Tan).*

**Abstract**

Liquid mobility on fibers is critical to the effectiveness of fiber matrices in face masks, water harvesting and aerosol filtration, but is typically affected by Plateau-Rayleigh instability. However, the spontaneous flow within precursor films arising from this instability has been largely overlooked, particularly regarding its fundamental flow pattern and the potential for liquid mobilization. This study reveals the pivotal role of spontaneous flow on ribbon-like fibers in enhancing liquid transport. The non-axisymmetric curvature of these fibers induces long-wave instabilities, generating a sustained flow that enables film-wise transport over centimeter-scale distances at velocities of several millimeters per second. Using particle-image velocimetry, we uncover intricate hydrodynamics, including opposing flows within the film and organized vortices in the shear layer, driven by capillary effects at the liquid-vapor interfaces. Building on these insights, we demonstrate a network structure capable of achieving planar liquid transport over a 10 cm$^2$ area. The ribbon-like fibers investigated exhibit the longest transport distances relative to biomimetic structures and aerodynamic propulsion. The unique transport dynamics and planar configuration of the fiber matrix offer substantial potential for advanced fiber-based liquid transport systems, with enhanced mass/heat transfer, laminar mixing and aerodynamic characteristics.

## 1. Introduction

The deposition of droplets on fiber matrix is a common yet critical process in various applications. Microfibers in face masks can effectively capture respiratory droplets from human breath, thereby deterring the transmission of bacteria and viruses present in saliva[1,2]. Large fabric nets are deployed in mountainous and coastal regions to harvest fog droplets as an alternative water source[3-5]. In industrial settings, fibrous media are used to filter oil mists generated during lubrication processes, mitigating environmental hazards[6,7]. In all these scenarios, the mobility of liquid along fibers and the dynamics of the mobilization are essential to the effectiveness of fiber matrices, particularly concerning liquid clogging and drainage[8-11].

However, hydrodynamic instabilities often arise, complicating the mobilization of liquid on fibers. Specifically, the liquid deposited on fiber surface typically forms a film that is prone to environmental perturbations[12]. Due to destabilizing capillary forces, these perturbations intensify, decomposing the film into liquid drops – a phenomenon known as Plateau Rayleigh Instability (PRI)[13]. Mobilizing the resulting drops (PRI drops) is challenging, because they face a threshold resistance caused by contact angle hysteresis[14].

Liquid mobilization along fibers can be achieved by directionally propelling PRI drops through aerodynamic effects[15,16] or complex geometrical configurations of fibers[17-19]. However, the PRI itself may also be harnessed for this purpose. This study is inspired by the observation that PRI drops are usually connected by a microscopic precursor film[20-22], enabling the deposited liquid to spontaneously flow towards nearby PRI drops[23]. Such spontaneous flow has long been observed on cylindrical fibers but shows limited potential for mobilizing liquid over long distances[23,24], because the fiber surfaces are predominantly covered by PRI drops[25]. Nevertheless, we speculate that long-distances liquid transport along fibers is possible if the PRI drops are spaced sufficiently apart.

Here we report on the liquid transport on ribbon-like fibers arising from PRI. We find that the asymmetric curvature of ribbon-like fibers drives the liquid away from the highly curved "shoulder" to the flatter "belly", triggering long-wave PRI on the belly side. This long-wave PRI results in spaced PRI drops, enabling the precursor film to transport liquid over centimeter-scale distances. Within the film-wise transport, we observe a pair of opposing flows across the fiber belly, which is narrower than 1 mm, moving at a few millimeters per second. The shear layer between the opposing flows undergoes non-linear development of Kelvin-Helmholtz instability, leading to the formation of organized micrometer-scale vortices. These phenomena are explained by thin-film hydrodynamics dominated by capillarity due to the shape of liquid-vapor interfaces. Finally, we apply the obtained knowledge to realize planar liquid transport across a fiber network. These findings have significant implications for designing fiber-based systems for liquid transport, offering promising technical applications.

## 2. Results

During liquid deposition on fibers, PRI develops from the fastest growing mode of environmental perturbations. As shown in Fig. 1a, the deposited liquid forms a thin liquid film with a thickness $h$ wrapping around the fiber. Environmental perturbations, existing as a series of sinusoidal modes, create crests and troughs on the liquid-vapor interfaces. The trough interface generates a greater capillary pressure in liquid phase than the crest interface, driving the liquid to flow towards the crest and eventually decomposing the liquid film into drops. Linear stability analysis shows that a single mode with wavelength $\lambda$ has the fastest growth rate, thus dominating the inter-drop distance[26]. For a cylindrical fiber with a radius $a$, a thin liquid film ($h \ll a$) develops PRI with the dominating wavelength $\lambda$ and characteristic growth time $\tau_p$ as follows:

$$\lambda = 2\pi\sqrt{2}a \tag{1}$$

$$\tau_p = \frac{12\mu a^4}{\gamma h^3} \tag{2}$$

where $\mu$ and $\gamma$ are the liquid viscosity and surface tension. Though limited to cylindrical fibers, Eqs. (1) and (2) reveal two characteristics of PRI: (a) the dominating wavelength is controlled by the surface curvature of the fiber, and (b) the thickness of film determines the growth rate of PRI.

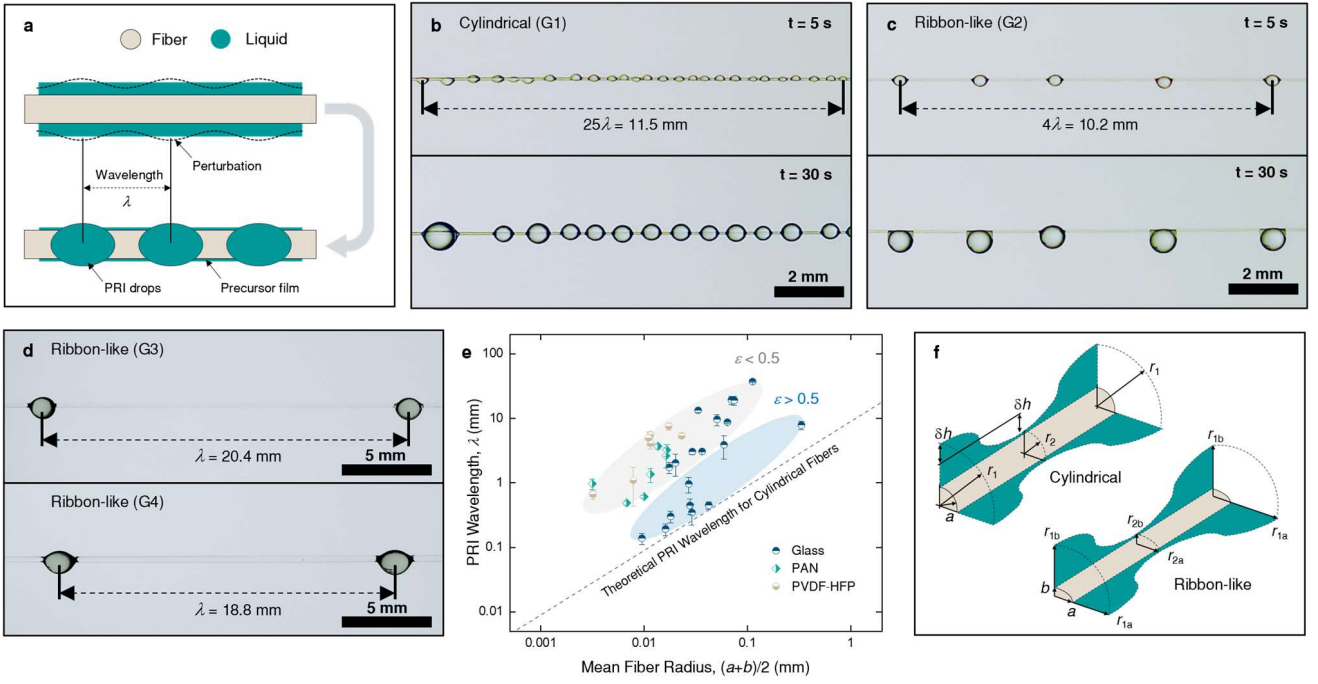

**Fig. 1: PRI on cylindrical and ribbon-like fibers during liquid deposition. a**, Schematic of PRI on a fiber developing from the perturbed liquid film to the decomposed liquid drops. **b-d**, Digital images of PRI drop formed when exposing glass fibers to liquid deposition at a flux of 14 μm s$^{-1}$. **b**, Cylindrical glass fiber (G1, $a$ = 41 μm) exposed for 5 seconds and 30 seconds; **c**, Ribbon-like glass fiber (G2, $a$ = 41 μm, $b$ = 18 μm) exposed for 5 seconds and 30 seconds. **d**, Ribbon-like glass fibers (G3, $a$ = 117 μm, $b$ = 23 μm; G4, $a$ = 160 μm, $b$ = 43 μm) exposed for 30 seconds. **e**, Plot of PRI wavelength as a function of mean fiber radius $(a+b)/2$. **f**, Schematic of perturbed liquid-vapor interface on the cylindrical and ribbon-like fiber. Data points and error bars in **e** indicate mean values ± standard deviation (s.d.) of four measurements.

Due to differences in surface curvature, cylindrical and ribbon-like fibers exhibit distinct PRI wavelengths. Fig. 1b shows the PRI drops on a cylindrical glass fiber (G1) with $a$ = 41 μm after 5 seconds and 30 seconds of exposure to liquid (deionized water droplets) deposition at a flux of 14 μm s$^{-1}$. The flux is the liquid flow (μm$^3$ s$^{-1}$) per unit area (μm$^2$) normal to the fiber. At 5 seconds, a liquid film with the length of 11.5 mm decomposes into 26 PRI drops, indicating a dominating wavelength of 0.46 ± 0.12 mm. This wavelength matches the theoretical value calculated using Eq. (1), but with an average increase of 26%. Similarly, Haefner et al. reported an average 28% increase in PRI wavelength on cylindrical glass fiber compared to their theoretical model[13], due to the experimental contribution of several perturbation modes. With continuous liquid deposition, the number of PRI drops reduces from 26 to 12 at 30 seconds due to drop growth and merger.

In contrast, PRI on ribbon-like fibers develops longer wavelengths, enabling spontaneous film-wise transport. We test a ribbon-like glass fiber (G2) with a major radius of $a$ = 41 μm and a minor radius of $b$ = 18 μm under the same conditions as G1. The aspect ratio of the fiber, $\varepsilon = b/a$ = 0.44, represents the flatness of fiber cross section. As shown in Fig. 1c, the dominating wavelength measured after 5 seconds of exposure is 2.55 ± 0.43 mm, which is 6 and 15 times longer than the theoretical wavelength calculated using the fiber radii $a$ and $b$, respectively. Despite the continuous liquid deposition, the number of drops remains unchanged after 30 seconds, indicating that PRI does not occur consecutively between two neighboring drops. Furthermore, the number of drops remains constant throughout the test until the drops grow sufficiently large to detach from the fiber. Along with the increasing drop volume, the constant number of drops evidences the sustained film-wise transport of liquid to the PRI drops over millimeter-scale distances.

Further achieving centimeter-scale film-wise transport and PRI wavelength is feasible by flattening the ribbon-like fibers, i.e., increasing the major radius and reducing the aspect ratio. As shown in Fig. 1d, we use the same condition to test two "flat" ribbon-like fibers G3 ($a$ = 117 μm, $b$ = 23 μm) and G4 ($a$ = 160 μm, $b$ = 43 μm), with aspect ratios of 0.2 and 0.27, respectively. The deposited liquid develops PRI of wavelength of 20.4 mm on G3 and 18.8 mm on G4. The inter-drop distance, determined by the wavelength, remains unchanged after 30 seconds, indicating centimeter-scale film-wise transport is sustained. The increased wavelength from millimeter to centimeter scale is primarily related to the major radius of G3 and G4 being 2.9 and 3.9 times greater than that of G2. Additionally, the aspect ratio affects the wavelength, because G3 with a smaller aspect ratio has a longer wavelength than G4, despite G4 having a larger major radius.

Further experiments validate the long-wave PRI characteristics of ribbon-like fibers across varied sizes and materials. Under the same conditions as G1-4 fibers, we tested fibers with sizes in the range of 10 μm < $a$ < 400 μm and 1 μm < $b$ < 265 μm, including materials of glass, polyacrylonitrile (PAN), and poly(vinylidene fluoride-*co*-hexafluoropropylene) (PVDF-HFP), each with distinct wettability. Fig. 1e summarizes the PRI wavelengths of different ribbon-like fibers based on their mean fiber radius ($a$ + $b$)/2 for comparison with cylindrical fibers. Regardless of fiber material, the PRI wavelengths for ribbon-like fibers are longer than the theoretical wavelength for cylindrical fibers, confirming long-wave characteristics. Importantly, the wavelength is related to the aspect ratio; fibers with $\varepsilon$ < 0.5 have longer wavelengths than those with $\varepsilon$ > 0.5. Therefore, the long-wave characteristics are attributed to the flat morphology of the fiber cross section rather than the specific radius or material wettability.

A plausible speculation is that the long-wave PRI and resulting film-wise transport are based on the non-axisymmetric curvature of ribbon-like fibers. As shown in Fig. 1f, for cylindrical fibers, the perturbation of liquid-vapor interface with a magnitude of δ$h$ is axisymmetric, leading to an even distribution of capillary pressure $p \sim \gamma/r$ along the fiber circumference. Consequently, the capillary pressure along the fiber axis is determined by the radius of curvature at the crest and the trough, i.e., $r_1$ = $a+h+\delta h$ and $r_2$= $a+h-\delta h$, indicating that the fiber radius, $a$, controls the PRI wavelength when $h$ << $a$. In contrast, the non-axisymmetric curvature of ribbon-like fibers results in an uneven circumferential distribution of capillary pressure, controlled by the radii of curvature of the film, i.e., $r_{1a}$, $r_{1b}$, $r_{2a}$ and $r_{2b}$. Therefore, understanding how curvature affects the PRI wavelength and the hydrodynamics of the film-wise transport is essential.

## 2.1 Mechanism of long-wave PRI on ribbon-like fibers

The non-axisymmetric curvature of ribbon-like fibers induces a circumferential pressure difference that varies the thickness of liquid film. We approximate the cross section of the ribbon-like fiber as an ellipse with major and minor semiaxis of $a$ and $b$ (Fig. 2a). The thin liquid film ($h \ll a$) wrapping around the ribbon-like fiber leads to an uneven distribution of capillary pressure: $p_1 = \gamma/(b^2/a)$ at the narrow side (shoulder) and $p_2 = \gamma/(a^2/b)$ at the broad side (belly). The circumferential pressure difference drives the liquid away from the shoulder, causing the liquid to aggregate on the belly. A similar phenomenon is observed in nature, where butterflies and moths, when feeding on liquid nectar, aggregate the liquid on the broad side of their flattened proboscises[27]. Therefore, the non-axisymmetric curvature drives circumferential capillary flow, resulting in a thicker belly film and a thinner shoulder film.

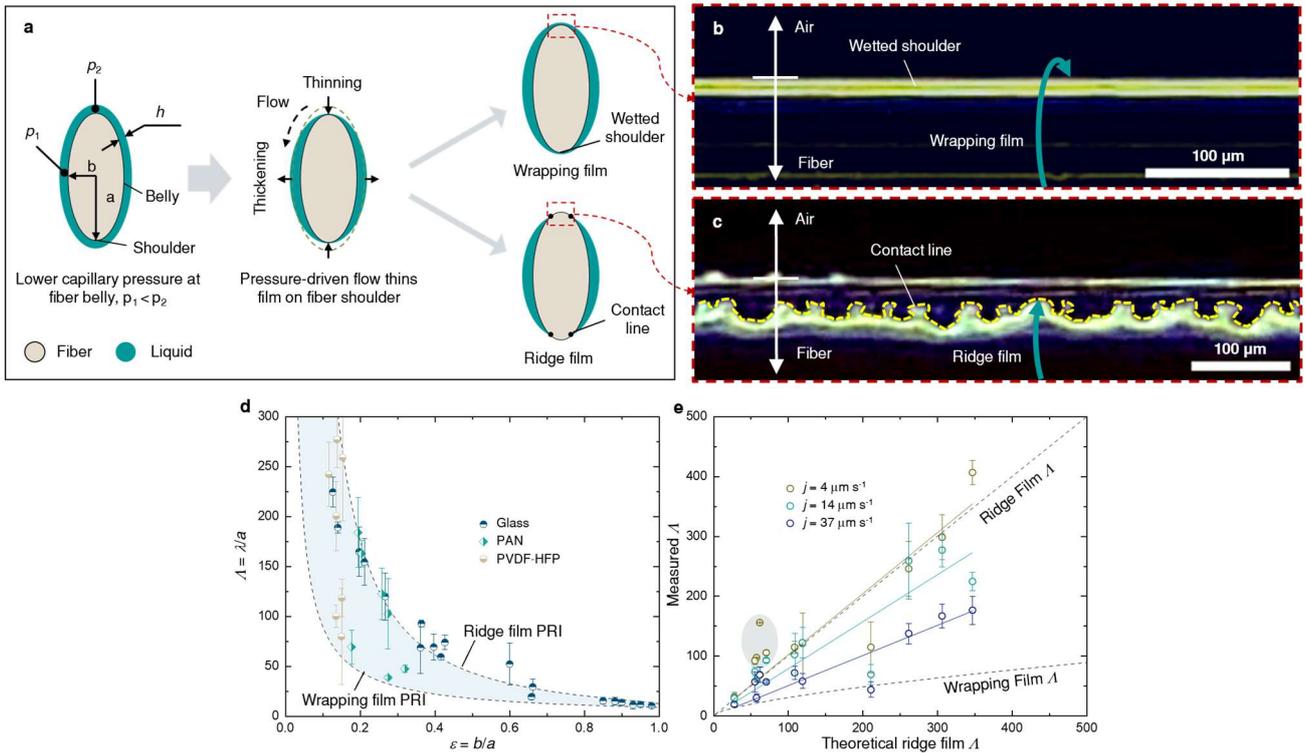

**Fig. 2: Development of long-wave PRI on ribbon-like fibers. a**, Schematic of the variation of film thickness along the circumference of the ribbon-like fibers and the resulting two possible film morphologies. **b, c**, Digital microscopic image of the film morphologies on the shoulder of ribbon-like glass fibers with the sizes of $a$ = 59 μm, $\varepsilon$ = 0.13 (**b**), and $a$ = 196 μm, $\varepsilon$ = 0.14 (**c**). **d**, Plot of the dimensionless PRI wavelength, $\Lambda$, as a function of fiber aspect ratio, $\varepsilon$, using the experimental data and the derived PRI wavelength for wrapping and ridge film. **e**, Plot of measured $\Lambda$ as a function of the theoretical $\Lambda$ for ridge film at the liquid flux, $j$, of 4 – 37 μm s$^{-1}$. The colored solid lines are the best linear fitting of the measured $\Lambda$. Data points and error bars in **d** and **e** indicate mean values ± s.d. of four measurements.

The variation in film thickness leads to two possible film morphologies: wrapping and ridge film. The wrapping morphology forms when the film remains intact and wets the fiber shoulder. As shown in Fig. 2b, the shoulder of a ribbon-like fiber ($a$ = 59 μm, $\varepsilon$ = 0.13) is wetted by the liquid, allowing the film to wrap around the fiber. In contrast, the thinned liquid film on the fiber shoulder ruptures when the capillary pressure difference overcomes the conjoining pressure between the liquid-vapor and liquid-solid interfaces[28]. As shown in Fig. 2c, the ruptured film retracts from the shoulder of a ribbon-like fiber ($a$ = 196 μm, $\varepsilon$ = 0.14), developing a dewetting pattern with a sinuous contact line due to the instability of the liquid rim[29]. Consequently, a liquid ridge forms on the fiber belly.

The two film morphologies further affect the onset of PRI. For the wrapping film, the thinning of the shoulder film increases the timescale for local perturbations growing into PRI due to $\tau_p \propto h^{-3}$ in Eq. (2), thus suppressing the onset of PRI on the fiber shoulder. In contrast, the belly film tends to rapidly develop PRI due to large thickness. For the ridge film, the shoulder film is ruptured and cannot be perturbed for PRI development. Therefore, PRI on ribbon-like fibers is dominated by the thickened liquid film on fiber belly.

The dominant role of fiber belly on PRI allows the theoretical derivation of wavelengths for wrapping and ridge films. The derivation (*see* Methods) is based on the principle that the wavelength is controlled by the surface curvature, avoiding the complicated description of the non-axisymmetric geometry. For the wrapping film, the theoretical PRI wavelength is derived from Eq. (1) for cylindrical fibers by replacing the fiber radius with an equivalent radius of the fiber belly:

$$\Lambda = \frac{\lambda}{a} = \frac{2\pi\sqrt{2}}{\varepsilon} \tag{3}$$

where $\Lambda$ is the dimensionless PRI wavelength. For the ridge film, perturbations need to move the contact line to deform the film, therefore viscous dissipation at moving contact lines should be included, leading to the following theoretical wavelength:

$$\Lambda = \frac{\lambda}{a} = \frac{2\pi}{\varepsilon}\sqrt{\frac{2\pi}{\arcsin \varepsilon}} \tag{4}$$

Eq. (3) and (4) suggests that the ridge film has a longer PRI wavelength than the wrapping film because $\arcsin\varepsilon < \pi$. The longer wavelength reveals the stabilizing effect of the contact line on short-wave PRI, where the perturbations are dampened by the viscous dissipation at the moving contact line. Additionally, as $\varepsilon$ approaches 0, both film morphologies yield an infinitely long wavelength, implying that the liquid film on an extremely flat ribbon-like fiber remains stable. This extreme case is surprisingly reasonable because the sharp shoulder with $\varepsilon = 0$ tends to pin the contact line, thus stabilizing the film from decomposing into drops, which aligns with the theoretical results of Davis[30].

The preceding theoretical PRI wavelengths are related to the aspect ratio, allowing convenient comparison with our experimental results. As shown in Fig. 2d, the wavelengths for most tested fibers fall within the region bounded by the curves of Eq. (3) and (4). Therefore, the long-wave PRI can be satisfactorily explained by our theoretical analysis. Additionally, the measured wavelengths for glass fibers collapse on the curve of the ridge film, whereas the polymeric fibers are scattered towards the curve of wrapping film. The scattered distribution is because polymeric fibers have rougher surface compared to the glass fibers (Supplementary Fig. 1), providing more surface for liquid attachment and preventing the rupture of the shoulder film[28]. Overall, non-axisymmetric ribbon-like fibers aggregate the deposited liquid on the fiber belly with low curvature, resulting in long-wave PRI.

Notably, the PRI wavelength varies with the flux of liquid deposition. As shown in Fig. 2e, the increasing liquid flux reduces the measured $\Lambda$ within the region bounded by the theoretical $\Lambda$ for the ridge film and wrapping film. At a liquid flux of $j = 4$ μm s$^{-1}$, the fitting line of the measured $\Lambda$ is closer to the theoretical value of the ridge film, whereas the line shift towards the wrapping film curve at a higher flux of $j = 37$ μm s$^{-1}$. The shift takes place because higher liquid flux can replenish the thinning film at the fiber shoulder, preventing the wrapping film from rupturing into a ridge film. In addition, the measured $\Lambda$ at $j = 4$ μm s$^{-1}$ systematically deviate from the bounded region when theoretical $\Lambda < 100$ (grey circle area). The deviation is attributed to the size difference of the PRI drops, where larger drops consume the smaller ones due to their different capillary pressure[31]. For two closely distanced drops with $\Lambda < 100$, the increased capillary suction amplifies the inter-drop consumption when a small flux $j = 4$ μm s$^{-1}$ cannot replenish the rapidly depleting liquid of smaller drops. As a result, the larger drop "eats" the smaller one, which appears as the increased wavelength.

## 2.2 Characteristics and hydrodynamics of liquid transport

After PRI occurs, the precursor film between PRI drops receives deposited liquid, evolving into a flowing film that transports the liquid to neighboring drops. As shown in Fig. 3a, the deposited liquid with high surface energy tends to curl into drops to reduce surface, generating a capillary pressure, $p_1$. In contrast, existing PRI drops with a radius $r$ have lower surface energy and smaller capillary pressure, $p_2$. The pressure difference $\Delta p = p_1 - p_2$ over a distance of half the PRI wavelength ($\lambda/2$) exerts a suction force on the liquid film, driving the flow towards the PRI drops.

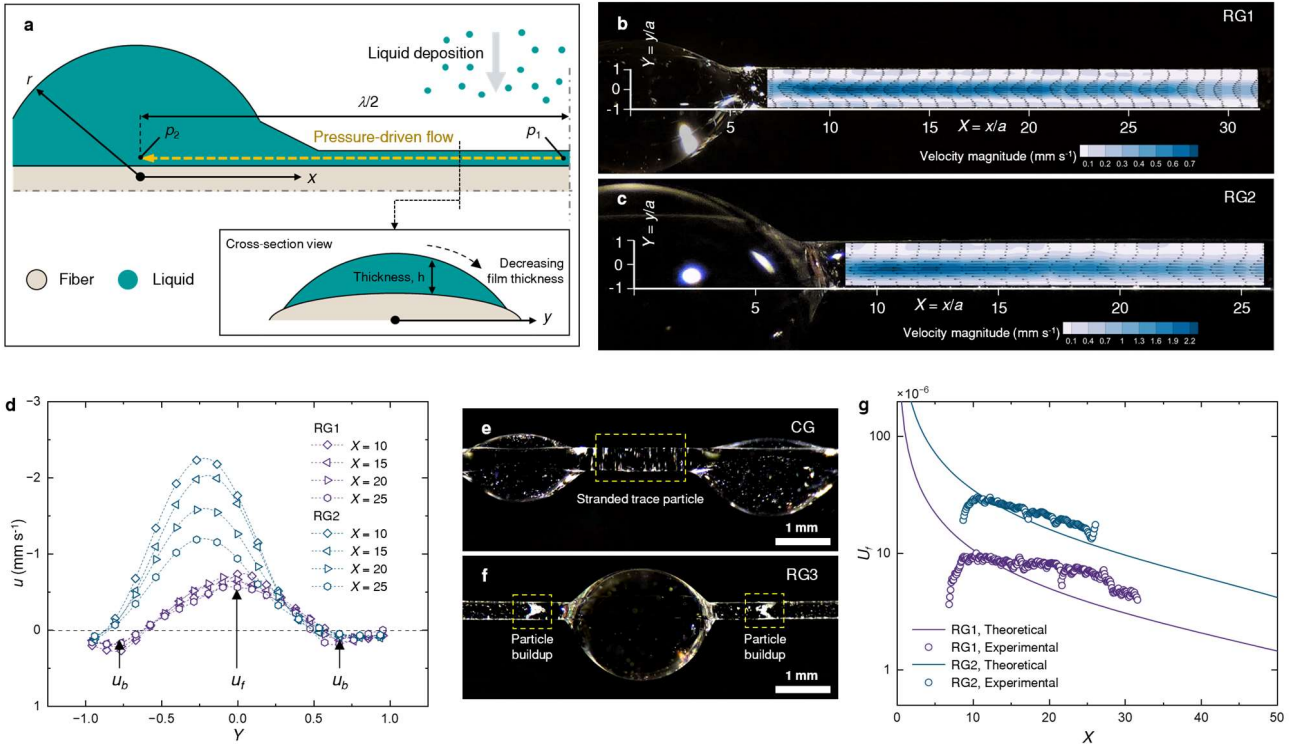

**Fig. 3: Overall flow field of the film-wise transport and formation of the forward flow. a**, Schematic of liquid transport from the liquid film to the PRI drop. The figure inset provides a cross-section view of the liquid film. **b, c**, PIV results of velocity vector (arrow) and magnitude (contour) for the film-wise flow on two ribbon-like fibers of RG1 ($a$ = 196 μm, $\varepsilon$ = 0.14, **b**) and RG2 ($a$ = 117 μm, $\varepsilon$ = 0.2, **c**). **d**, Plot of the $X$ component of the flow velocity, $u$, as a function of $Y$. **e, f**, Digital microscopic images of trace particles on CG (cylindrical glass fiber, $a$ = 125 μm, **e**) and RG3 (ribbon-like glass fiber, $a$ = 94 μm, $\varepsilon$ = 0.36, **f**). **g**, Plot of the dimensionless forward velocity, $U_f$, of RG1 and RG2 as a function of $X$. $U_f$ obtained from the theoretical model and the experimental measurements by PIV are provided for comparison.

Particle image velocimetry (PIV) measurement of two ribbon-like glass fibers, RG1 ($a$ = 196 μm, $\varepsilon$ = 0.14) and RG2 ($a$ = 117 μm, $\varepsilon$ = 0.2), reveal that the film-wise transport consists of a pair of opposing flows. PRI on RG1 and RG2 exhibit wavelengths of $\lambda$ = 37 mm and 19 mm, forming PRI drops with different radius (RG1: $r_1$ = 0.56 mm, $r_2$ = 0.96 mm; RG2: $r_1$ = 0.70 mm, and $r_2$ = 0.83 mm; 1 and 2 denote principle radii). The vector field and velocity magnitude of film-wise flow (Fig. 3b and c) are presented using dimensionless coordinates $X = x/a$ and $Y = y/a$. The flow velocity is in the range of 0 – 0.7 mm s$^{-1}$ for RG1 and 0 – 2.2 mm s$^{-1}$ for RG2. The liquid near the fiber belly flows towards the PRI drops at higher velocities (forward flow), whereas the liquid near the fiber shoulder flows away from the PRI drops at lower velocities (backward flow), resulting in recirculation.

Fig. 3d shows the profile of the $X$ component of the velocity, $u$, across the $Y$ axis at $X$ = 10, 15, 20, 25. Using RG1 as an example, the forward flow maintains a negative $u$ within 0.6 < $Y$ < 0.6, peaking at $u_f$ = -0.5 mm s$^{-1}$ – -0.75 mm s$^{-1}$ around $Y$ = 0, forming a symmetric Poiseuille-flow profile. This profile arises from the circular shape of the film cross-section (inset of Fig. 3a), with the film being thickest at $Y \approx 0$ and thinner towards the fiber shoulder. The thin shoulder film is subjected to the viscous boundary layer at the liquid-solid interface, which reduces flow velocity. The backward flow occurs when $|Y| > 0.6$, peaking at $u_b \approx 0.1$ mm s$^{-1}$ – 0.3 mm s$^{-1}$ at $|Y| \approx 0.7$. Additionally, the $u$ profile for RG2 shifts towards the lower shoulder ($Y = -1$) because the mass center of the heavy PRI drop gravitationally sags below the centerline of RG2, thus lowering the suction position.

We also seed the liquid film on a cylindrical fiber (CG, $a$ = 125 μm) and a rounder ribbon-like fiber (RG3, $a$ = 94 μm, $\varepsilon$ = 0.36) to visualize the flow. However, the liquid film on CG is too thin to drift the 10 μm diameter trace particles, resulting in "stranding" of the particles (Fig. 3e). Similarly, RG3 generates a thinner film compared to RG1 and RG2, causing particle buildup near the PRI drop (Fig. 3f), which artificially affects the flow field. The particle stranding implies that the film thickness is closely related to the aspect ratio.

The mechanisms and dynamics of the opposing flows require further understanding. Knowing that the forward flow is generated by the capillary pressure difference, we model the forward flow (*see* Methods) by casting variables into dimensionless form using the major radius, $a$, and capillary velocity, $\gamma/\mu$. The dimensionless velocity of forward flow $U_f = |u_f \mu/\gamma|$ is given by:

$$U_f = \frac{2H^2}{3}\left(H + \varepsilon - \frac{1}{R}\right)\left(\frac{1}{X} - \frac{1}{\Lambda - X}\right) \tag{5}$$

where $H = h/a$, $R = r/a$, and $r$ is the equivalent radius of the PRI drop by curvature $r = 2/(1/r_1 + 1/r_2)$. The film thickness, $H$, is obtained by applying conservation between the flux of the deposited liquid ($J = j\mu/\gamma$) and the forward flow.

$$J = \frac{4H^3}{3\Lambda R}\left(H + \varepsilon - \frac{1}{R}\right) \tag{6}$$

In Eq. (5), $H+\varepsilon-1/R$ represents the contribution of capillary pressure difference due to the film curvature, $2H+2\varepsilon$, and the drop curvature, $2/R$. Additionally, $1/X-1/(\Lambda-X)$ represents the contribution of the distance from liquid film to two neighboring PRI drops. Together, these terms indicate that neighboring PRI drops compete for the deposited liquid over the distance of the PRI wavelength. Moreover, Eq. (6) indicates that, at a fixed liquid flux $J$, a rounder ribbon-like fiber with a larger $\varepsilon$ and smaller $\Lambda$ has a smaller film thickness $H$ compared to flatter fibers, explaining the particle stranding observed for fibers CG and RG3.

The forward flow model is validated by PIV results. As shown in Fig. 3g, the measured $U_f$ for RG1 and RG2 aligns well with the theoretical model at $X > 10$, with $R^2 \approx 0.68$ and 0.91, respectively. The decrease in $U_f$ at larger $X$ reflects a reduced net suction force due to a longer suction distance between the competing PRI drops and increased viscous dissipation over this distance. However, the measured $U_f$ deviates from the model when approaching the PRI drop ($X < 10$) due to the film-drop transition. In the transition zone, the film thickness increases to the scale of the PRI drop, enlarging the flow area. Consequently, $U_f$ for RG1 and RG2 decreases to 40-60% of the peak value. Overall, the model effectively explains the dynamics of forward flow at a sufficient distance from the PRI drop.

Fig. 4a explains the formation of the backward flow. Due to strong suction at adjacent distances, the highly curved shoulder film in the transition zone consistently squeezing the liquid into the PRI drop, forming an ultra-thin squeezing film (Fig. 4b). The small thickness of the squeezing film decreases the flow velocity since $u \propto h^2$, which increases the Bernoulli pressure of the film by $\Delta p \sim \rho \Delta(u^2)$. The resulting adverse pressure gradient against the nearby forward flow develops a separation front near the boundary of the squeezing film, reversing the direction of flow and providing a mass source for the backward flow (Fig. 4c). Additionally, the loss of liquid mass on the fiber shoulder due to viscous entrainment into the forward flow leads to the thinning of the shoulder film. Consequently, the reduced capillary pressure due to film thinning builds a pressure gradient along the fiber axis, creating a reverse suction force to replenish the lost liquid mass. Essentially, the squeezing film cuts off the pressure "communication" between the film and the PRI drop, amplifying the reverse suction to drive the backward flow.

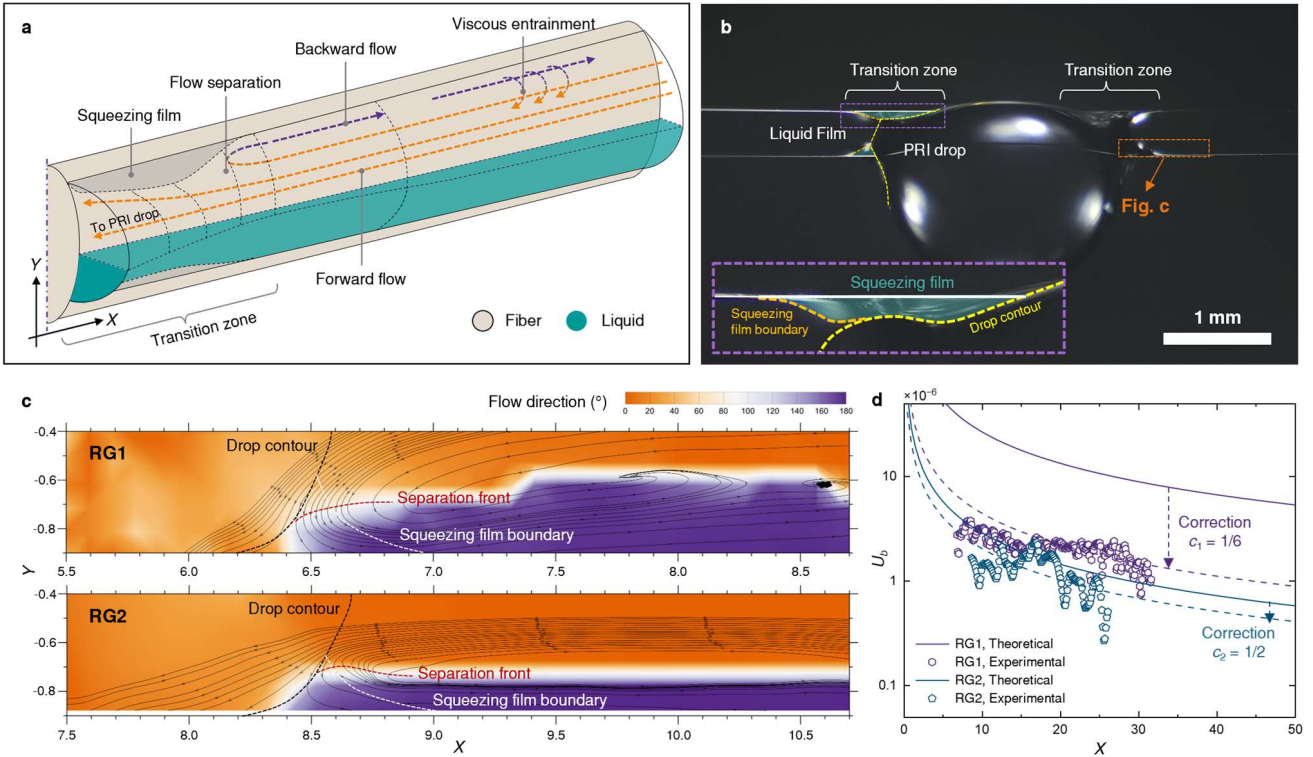

**Fig. 4: Formation of the backward flow. a**, Schematic of formation of the backward flow. **b**, Digital microscopic image of the squeezing film formed near fiber shoulder adjacent to the PRI drop. **c**, PIV results of flow direction (contour) and streamlines (arrowed lines) for the film-wise flow in the region enclosed by orange dashed box in **b**. **d**, Plot of the dimensionless backward velocity, $U_b$, for RG1 and RG2 as a function of $X$. $U_b$ obtained from the theoretical model and the experimental measurements by PIV are provided for comparison.

We modeled the backward flow based on the preceding mechanism by scaling the pressure gradient, leading to the following the dimensionless velocity of backward flow (*see* Methods)

$$U_b = \frac{H^3}{12X} \tag{7}$$

Here, $H$ is the thickness of liquid film at the fiber belly, which is calculated by Eq. (6). As shown in Fig. 4d, the model predicts $U_b$ with the same trend and order of magnitude as the PIV measured results. However, the model overestimates $U_b$ because the scaling of pressure gradient overestimates the liquid loss due to viscous entrainment. Particularly, the value of $U_b$ for RG1 is more deviated compared to RG2, because the forward velocity $U_f$ for RG1 is smaller compared to RG2, leading to less entrained liquid mass and larger overestimation. Introducing correction factors (RG1: $c_1 = 1/6$; RG2: $c_2 = 1/2$) to the scaled pressure gradient causes the experimental data to collapse on the model curves. Notably, $c_1$ is 3 times smaller than $c_2$, which coincides with the ratio between the values of $U_f$ for RG1 and RG2, implying a proportional relationship between the forward velocity and viscous entrainment. In addition, the measured $U_b$ fluctuates because of the vortices formed in the shear layer between the opposing flows, which is discussed later. Overall, the model explains the dynamics of backward flow and predicts the upper bound of $U_b$.

Using the models for the opposing flows, we can further discuss the characteristics of the film-wise transport. Eq. (6) indicates that, at a constant liquid flux $J$, the thickness $H$ varies according to the aspect ratio $\varepsilon$ and the PRI drop radius $R$. Specifically, for a flat ribbon-like fiber with small $\varepsilon$, the film thickness $H$ increases to overwhelm the contribution of $\varepsilon$. Meanwhile, the term $1/R$ usually vanishes because the drop radius, $r$, can grow much larger than $a$ ($R \gg 1$) due to the continuous transport. This useful scenario simplifies Eq. (6) as follows:

$$H \approx \left(\frac{3}{4} J \Lambda R\right)^{\frac{1}{4}} \tag{8}$$

Therefore, Eq. (5) and (7) can be simplified as

$$U_f \approx 0.537 (J \Lambda R)^{\frac{3}{4}} \left(\frac{1}{X} - \frac{1}{\Lambda - X}\right) \tag{9}$$

$$U_b \approx 0.067 \frac{(J \Lambda R)^{\frac{3}{4}}}{X} \tag{10}$$

Eq. (9) indicates a linear relationship between $U_f$ and $1/X$ when $X \ll \Lambda\text{-}X$, which is meaningful for long-wave PRI. Therefore,

$$U_f \approx 0.537 \frac{(J \Lambda R)^{\frac{3}{4}}}{X} \tag{11}$$

Eq. (10) and (11) reveal two characteristics of the transport. First, both the forward and backward velocity scales with $(J\Lambda R)^{3/4}/X$. Second, the forward velocity is 8 times greater than the upper bound of the backward velocity, thus governing the film-wise transport.

The scaling relation of $U_f$ can be generalized to determine transport capacity. In Fig. 5a, the forward velocity for RG2 varies linearly with $(J\Lambda R)^{3/4}/X$ within the range of 2.5 – 5.5, aligning with the scaling relation. In contrast, RG1 deviates because its PRI drop radius ($R = 3.6$) is insufficiently large to meet the criterion of $R \gg 1$. However, PRI drops inevitably grow during long-term transport, making the scaling practically robust. Additionally, the obedience of $U_f$ to the scaling relation divides the film-wise transport into transition, linear, and competing phases. In the transition phase, $U_f$ scales negatively with $(J\Lambda R)^{3/4}/X$ due to film-drop transition. In the competing phases, $U_f$ is non-linear, reaching $U_f = 0$ when $X = \Lambda/2$ (Supplementary Fig. 2) due to the rising contribution of $1/(\Lambda-X)$ in Eq. (9). Clearly, the transport capacity is determined by the linear phase flow where $Q_f = 2U_fH = J\Lambda R/X$. Therefore, ribbon-like fibers with long-wave PRI improve the transport capacity by providing a long linear phase.

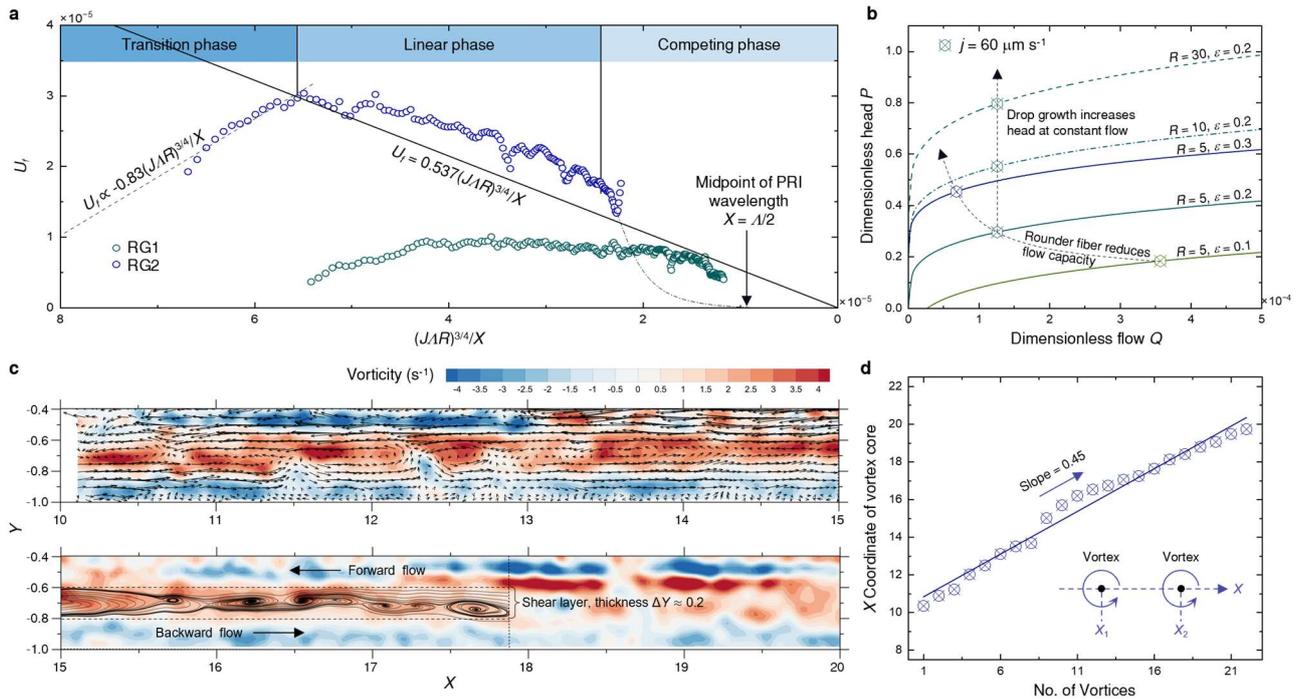

**Fig. 5: Characteristics of film-wise liquid transport and Plateau-Rayleigh pump**. **a**, Plot of the scaling relation between $U_f$ and $(J\Lambda R)^{3/4}/X$ for three different phases of forward-flow transport. **b**, Plot of characteristic (flow-head) curves of the Plateau-Rayleigh pump at $R = 5, 10, 30$ and $\varepsilon = 0.1, 0.2, 0.3$. **c**, PIV results of the vorticity (contour), velocity vector (arrows) and streamline (lines) in the shear layer between the opposing flows for RG1. **d**, Plot of the $X$ coordinates of shear-induced vortex cores as a function of the number of vortices. The solid line in **d** is the best linear fitting with slope = 0.45.

The film-wise transport on ribbon-like fibers can be compared to mechanical pumps. The fiber-film-drop system forms a "Plateau-Rayleigh pump" (PR pump): the fiber receives deposited liquid as the intake; the film generates capillary head as the impeller; the drop serves as the outlet to a liquid reservoir. Using our model, the dimensionless flow ($Q$) and head ($P$) of the PR pump are $Q = Q_f|_{X=R} = J\Lambda$ and $P = 2H+2\varepsilon-2/R$, respectively. Using Eqs. (4) and (8) to replace $\Lambda$ and $H$ with measurable $\varepsilon$, $J$, and $R$, the pump flow monotonically increases with head at a fixed $\varepsilon$ of 0.2 (Fig. 5b). As the radius $R$ increases from 5 to 30 during drop growth, the head rises at a constant flow capacity and liquid flux (dots show $j = 60$ μm s$^{-1}$). Similarly, increasing $\varepsilon$ from 0.1 to 0.3 at a constant $R = 5$ also raises the pump head. However, at a constant liquid flux, rounder fibers ($\varepsilon = 0.2$ and 0.3) exhibit reduced flow capacity compared to flatter fiber ($\varepsilon = 0.1$). Overall, flat fibers generally benefit situations that require high flowrates, while maintaining a large drop favors situations where a high pressure head is needed.

Interestingly, a shear layer forms between the opposing flows due to velocity discontinuity, triggering Kelvin-Helmholtz instability (KHI) that rolls up micrometer-scale vortices (Fig. 5c). Specifically, in a zoomed region ($15 < X < 18$) of RG1 containing the shear layer, eight organized elliptical vortices rotate counter-clockwise with a vorticity around 4 s$^{-1}$ in the shear layer with a dimensionless thickness of $L = \Delta Y \approx 0.2$. These vortices result from small perturbations rolling up the interface of velocity discontinuity and growing into KHI. Consequently, the vortices are separated by the distance of the dominating KHI wavelength, which can be predicted as $\Lambda_{KHI} \approx 2.5\pi L = 1.57$ using linear stability analysis[32]. However, the measured inter-vortex spacing averages 0.45 (Fig. 5d), 71% smaller than predicted. This deviation may be caused by the non-linear development of KHI, where sub-harmonics reduce the wavelength[33]. The micrometer-scale vortices enlarge interfaces between fluid blobs, thereby enhancing mixing of the liquid on fibers[34], which is normally difficult due to the laminar characteristics of the low-Reynolds film flow[35] ($Re < 1$ for RG1 and 2).

## 2.3 Planar liquid transport on fiber networks

A single ribbon-like fiber effectively transports liquid, but practical situations typically involve fiber networks. A network of ribbon-like fibers inevitably introduces intersections of fibers, which creates lower-energy surfaces that leads to the drop formation and disrupts liquid transport. To achieve planar liquid transport over a fiber network, we design a rivulets-on-web (ROW) structure to transport liquid continuously without forming drops at fiber intersections. As shown in Fig. 6a, a web consisting of vertical and horizontal fibers is woven with proper spacing. The horizontal fibers are ribbon-like, preventing the formation of PRI drops over the spacing $d_1 < \lambda$ due to long-wave PRI. The vertical fibers are arranged so that deposited liquid between two closely spaced fibers separated by $d_2$ forms a stable liquid bridge, whereas more distant fibers separated by $d_1$- $d_2$ cannot. This configuration uses the stable liquid bridge as an artificial PRI "drop". Consequently, the ribbon-like fibers can spontaneously transport the deposited liquid as "branch rivulet" into the liquid bridge, which acts as a "main rivulet" to further transport.

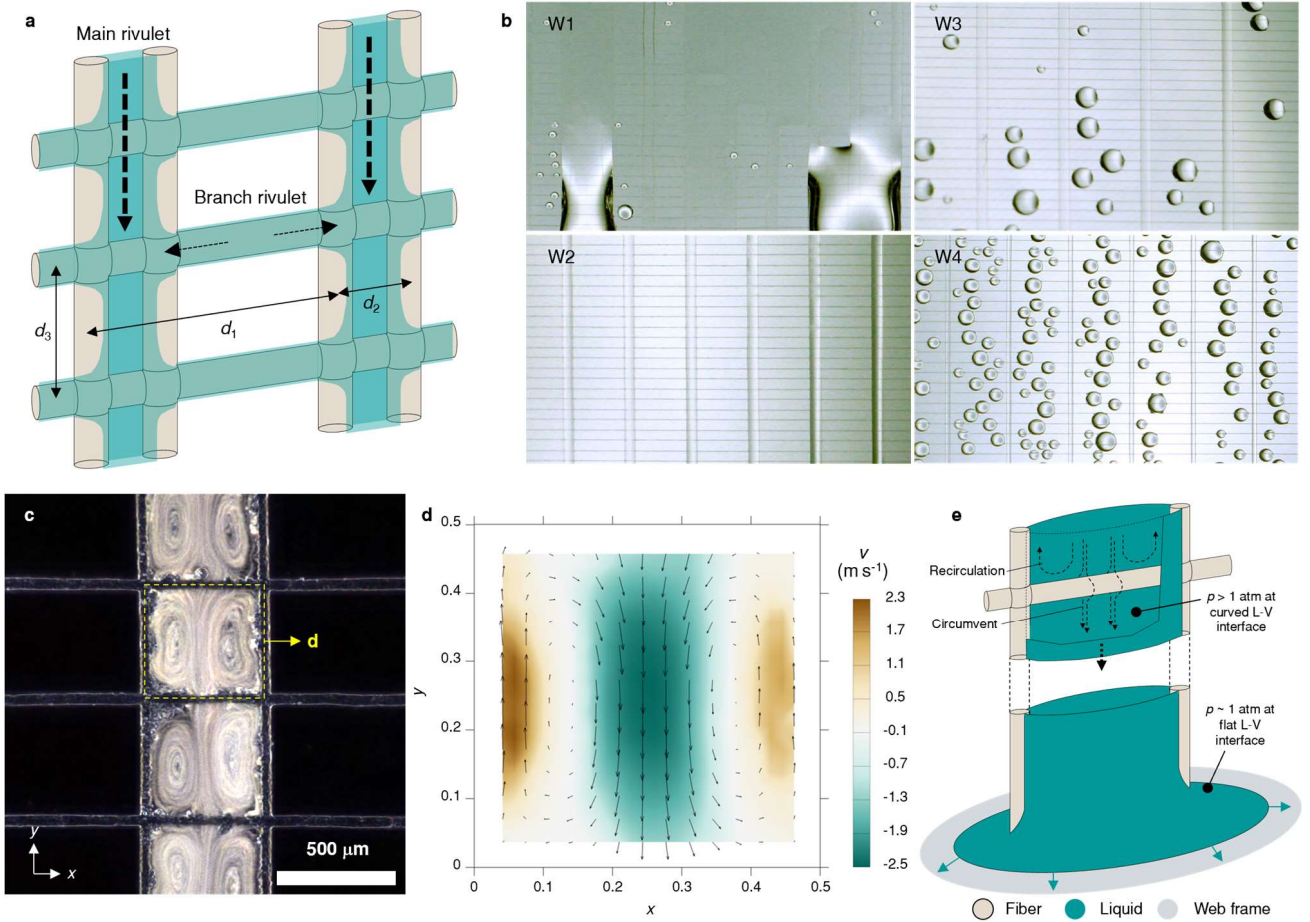

**Fig. 6: Liquid transport on rivulets-on-web (ROW) structures. a**, Schematic of conceptual design for ROW structure. **b**, Digital image of four different fiber webs (W1-W4) exposed to liquid deposition at a flux of 37 μm s$^{-1}$ for 30 minutes. **c**, Digital image of particle trajectory along the liquid transport in the main rivulet. The trajectory image is obtained by time projection of 500 consecutive video frames. **d**, PIV results of vertical component of velocity (contour) and velocity vector (arrow) for the main rivulet in the region enclosed by the yellow dashed box in **c**. **e**, Schematic of the capillary pressure gradient and vortex formation in the main rivulet.

Based on the conceptual design, four fiber webs (W1 – W4, Table 1) are fabricated using near-field electrospinning and fixed with two copper gaskets as the frame (Supplementary Fig. 3). Each fiber web, covering an area of 10 cm$^2$, consists of PVDF-HFP ribbon-like fibers with $a$ = 40 μm and $\varepsilon$ = 0.14. The spacing $d_1$ is controlled within the range of 2 mm – 4 mm, aligning with the dominating PRI wavelength of 1.3 mm (wrapping film) – 6 mm (ridge film). To explore the impact of main rivulet width on the ROW structure, two different $d_2$ of 0.2 mm and 0.5 mm are used. Additionally, $d_3$ is kept constant at 0.5 mm because it affects the ROW structure by influencing the stability of the liquid bridge, which is beyond the scope of this study.

**Table 1: Geometrical parameters for the fabricated fiber webs**

|  | $a$ (μm) | $\varepsilon$ | $d_1$ (mm) | $d_2$ (mm) | $d_3$ (mm) |
|---|---|---|---|---|---|
| W1 | 40 | 0.14 | 2 | 0.5 | 0.5 |
| W2 |  |  | 3 | 0.5 | 0.5 |
| W3 | | | 4 | 0.5 | 0.5 |
| W4 | | | 3 | 0.2 | 0.5 |

The fabricated fiber webs are exposed to liquid deposition at fluxes, $j$, in the range of 4 – 37 µm s$^{-1}$ to form the ROW structure. Within 30 minutes, the tested webs are covered with liquid bridges (Fig. 6b). For W1, the deposited liquid overwhelms the spacing $d_1$ at the lowest flux ($j = 4$ µm s$^{-1}$), spreading the liquid bridge across the web. In contrast, W2 isolates the liquid bridge into vertical main rivulets and prevents the formation of PRI drops even at the highest flux ($j = 37$ µm s$^{-1}$), indicating a successful ROW structure. W3 also isolates the liquid bridge but forms PRI drops between the main rivulets at $j = 37$ µm s$^{-1}$. W4, with a smaller $d_2 = 0.2$ mm, exhibits a similar behavior to W3, because the narrow main rivulets cannot provide sufficient suction on branch rivulets, triggering PRI on horizontal fibers. Upon the formation of ROW structure, W2 maintains continuous liquid transport throughout a 3-hour test at the highest flux of 37 µm s$^{-1}$ without drop formation.

To understand the transport characteristics of the ROW structure, we visualize the flow in the main rivulets of W2 using trace particles. At a liquid flux of $j = 4$ µm s$^{-1}$, the particle trajectory (Fig. 6c) shows downward flow near the rivulet centerline, flanked by a pair of counter-rotating vortices within each mesh. The PIV analysis (Fig. 6d) indicates the maximum downward velocity is 2.4 mm s$^{-1}$ at the mesh center. At a higher flux of $j = 37$ µm s$^{-1}$, the maximum velocity increases to about 3.8 mm s$^{-1}$, although the accurate PIV measurement is not obtained due to the foggy field of view at a higher flux. Therefore, the main rivulet thickness increases 5.8 times to transport more liquid, as velocity only increases 1.6 times while the liquid flux rises 9.25 times.

The downward flow is driven by the capillary pressure gradient along the main rivulet, which results from the curvature difference of the liquid-vapor interface. As shown in Fig. 6e, the liquid-vapor interface between two fibers has positive curvature due to surface tension, creating a capillary pressure larger than the ambient pressure. However, the interface is flattened when the liquid spreads on the wettable web frame, reducing the pressure to about 1 atm. This pressure gradient drives the liquid downward to the web frame. Additionally, vortex formation occurs due to the constrained flow around the horizontal fiber. The positively curved liquid-vapor interface near the rivulet centerline allows the downward flow to circumvent the horizontal fiber. Conversely, the film near the rivulet edge is sufficiently thin, leading to obstruction and flow separation around the horizontal fiber, resulting in recirculation. Overall, the ROW structure transports liquid to wettable surfaces by adjusting the thickness of the main rivulets, while forming symmetric vortex structures along the transport path.

## 3. Discussion

The liquid transport arising from PRI demonstrates exceptional transport distances compared to other approaches. As shown in Fig. 7a, various strategies for liquid transport along one-dimensional slender bodies are summarized based on their transport velocities and distances. The type of transport – whether drop-wise or film-wise – and the nature of deposition, ranging from micrometer-scale droplets to millimeter-scale bulk drops, are provided for comparison. Notably, the Plateau-Rayleigh pump in this work achieves the longest transport distances, ranging from 19 to 37 mm, with a moderate transport velocity of approximately 1 mm s$^{-1}$. In contrast, natural and biomimetic structures, despite utilizing complex geometries to enhance liquid transport, exhibit shorter transport distances. While aerodynamic propulsion also achieves long transport distances for bulk drops, it necessitates airflow to generate wake interactions[15,16]. Therefore, this study provides a simple approach to achieving long-distance liquid transport without complex modification of fiber geometry or introducing airflow.

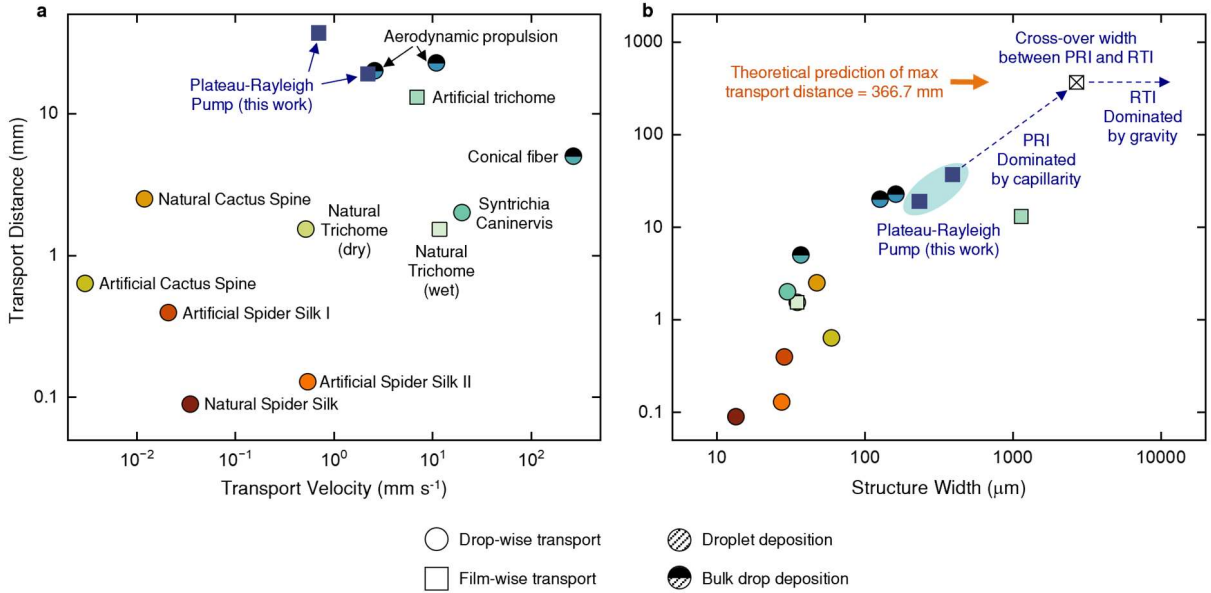

**Fig. 7: Comparison of transport distance between the Plateau-Rayleigh pump and other methods of liquid transport on slender bodies. a,** Plot of transport distance versus transport velocity. **b**, Plot of transport distance versus structure width. Data are obtained from references[15-17,36-41].

In addition, the transport distance of the Plateau-Rayleigh pump can be further enhanced by increasing the width (twice the long radius $a$) of the ribbon-like fibers. As shown in Fig. 7b, with a practically feasible aspect ratio of $\varepsilon = 0.15$, increasing the fiber width to the capillary length of water, $l_c = (\gamma/\rho g)^{1/2} = 2.71$ mm, leads to a theoretical maximum transport distance of 366.7 mm. Wider ribbon-like fibers introduce another type of instability dominated by gravity, i.e., Rayleigh-Taylor instability (RTI), which determine the inter-drop distances. The critical fiber width for instability crossover is approximately equal to $l_c$[42]. The liquid transport arising from RTI presents an intriguing direction for future investigation.

The findings of this work have significant technical implications. For example, film-wise transport can accelerate the renewal of liquid-vapor interface, reducing the local level of temperature or mass concentrations, thus enhancing heat and mass transfer. Additionally, the micrometer-scale vortices generated within the flow improve laminar mixing at low Reynolds numbers, offering potential for fiber-based designs of mixers or chemical reactors for low-*Re* applications. Furthermore, the suppression of drop formation by the ROW structure can improve the aerodynamic properties of fibrous media by reducing both pressure drop and droplet entrainment. Overall, this study offers new insights for designing fiber-based liquid transport systems with enhanced functional characteristics.

## 4. Methods

### 4.1 Preparation of cylindrical and ribbon-like glass fibers

Cylindrical glass fibers are prepared by pulling heated glass capillary tubes using a butane torch (Cole-Parmer, Canada). For Ribbon-like glass fibers, the heated glass capillary tubes are first pressed by a tweezer and then pulled. The size of fibers is measured by visualizing the cross section of fibers with microscope (AmScope SW-3T24Z, USA).

### 4.2 Preparation of ribbon-like polymeric fibers and fiber webs

Polymeric fibers and fiber webs are prepared using a near-field electrospinning (NFES) system. 2.5 g poly(vinylidene fluoride-*co*-hexafluoropropylene) (PVDF-HFP, Mw ~ 455,000, Sigma Aldrich, Canada) is dissolved into a mixture solvent of 3.75 g *N,N*-dimethylacetamide (DMAc, purity > 99.8%, Sigma Aldrich, Canada) and 3.75 g acetone (purity ≥99.5 %, Fisher Scientific, Canada) to prepare a PVDF-HFP solution of 25 wt% concentration. 1 g polyacrylonitrile (PAN, Mw ~ 150,000, Sigma Aldrich, Canada) is dissolved into a mixture solvent of 4 g dimethyl sulfoxide (DMSO, purity > 99.9%, Sigma Aldrich, Canada) and 1.55 g acetone to prepare a PAN solution of 18 wt% concentration. The solutions are stirred overnight at room temperature.

The prepared polymer solutions are loaded into a glass syringe (5 ml, Hamilton, USA). The syringe is then connected to a 26 gauge metallic needle (Sigma Aldrich, Canada) and installed into the NFES system. The NFES system, which has been described in our previous work[43], deposits polymer solutions on a grounded collector at an electric voltage of 1.6 – 2 kV and a needle-to-collector distance of 1.5 – 2.5 mm. The collector translationally moves along a programmed path to deposit polymer solutions, which further solidates as single fibers or patterned fiber webs. The deposited wet solution on the flat collector spreads laterally, thus forming a ribbon-like morphology during solidification. The size of the ribbon-like fibers, i.e., *a* and *b*, are controlled by tuning the flow rate of polymeric solution in the range of 0.05 ml h$^{-1}$ – 0.2 ml h$^{-1}$ and the collector moving speed in the range of 25 mm s$^{-1}$ – 150 mm s$^{-1}$. The obtained ribbon-like fibers have the size in the range of 10 μm < $a$ < 45 μm and 1 μm < $b$ < 6 μm. The size of the fibers is measured by visualizing the cross section of fibers with scanning electron microscope (Zeiss FESEM 1530, Germany).

## 4.3 Experimental Setup

Experiments are conducted using a homemade wind tunnel (Supplementary Fig. 4). An axial fan (Noctua NV-FS2, Austria) is installed at the inlet of the tunnel to generate airflow. The fan speed is controlled by a DC power supply that regulates voltage from 5 V to 12 V. The airflow is conditioned using a tube-bundle flow straightener to eliminate the swirls generated by the axial fan. An ultrasonic humidifier injects droplets of deionized water into the airflow, providing a continuous supply of liquid with a tunable flowrate in the range of 18 – 167 ml h$^{-1}$, leading to a flux of 4 – 37 μm s$^{-1}$ at the outlet of the wind tunnel. The droplet diameter is in the range of 2 – 50 μm with a mean diameter of 16 μm, which is sufficiently small compared with the fiber diameter to avoid liquid splashes created by droplet-fiber collisions. Ducts and contractions are used to connect the above components to complete the wind tunnel. The airflow velocity inside the tunnel can be adjusted within the range of 1.4 – 2.2 m s$^{-1}$ by controlling the fan speed; however, the velocity is kept as 2 m s$^{-1}$ throughout this work because it imposes negligible influence on test results. A digital camera (Cannon R5, Japan) attached to a microscope (AmScope SW-3T24Z, USA) is placed at the rear side of the tunnel for visualization.

The PRI wavelength on fibers is measured as the distance between two neighboring PRI drops. We place the tested fiber at the outlet of the wind tunnel so that liquid droplets can continuously deposit on fibers. Before each test, the fiber is immersed in the droplet-laden airflow for at least 1 minute to ensure an even coating of liquid film. Then, any formed drops on the fiber are removed by a gentle airflow, leaving a residual liquid film on the tested fibers. During the test, the liquid film gradually grows until PRI occurs and decomposes into drops. The microscopic camera records the image when the number of PRI drops stops increasing. The recorded images are processed to measure the inter-drop distance as the PRI wavelength.

The liquid transport on ribbon-like fibers is characterized by particle-image velocimetry (PIV). The tested ribbon-like glass fibers are put in a droplet flux of 14 μm s$^{-1}$ to trigger PRI onset, and the flux is reduced to 4 μm s$^{-1}$ for flow seeding with better image quality. The transparent glass fiber allows the camera to record the film flow on the rear side that receives droplet flux. We use polystyrene microspheres with diameters of 10 μm as trace particles to visualize the flow. The particles are screened using a 53 μm sieve (W.S. Tyler, Canada) to remove large clumps, and then delivered to the inlet of wind tunnel by compressed gas. The airflow inside the tunnel carries the particles to impact the tested fiber, thus seeding the liquid film. A fiber-optic lamp (Cole-Parmer, Canada) is placed parallel with the objective lens of the microscope to illuminate the particles. After the delivery of particles, the microscopic camera records videos of film flow on tested fiber at a frame rate of 120 fps. Notably, the liquid film is sparsely seeded with a number density of 200 – 300 mm$^{-2}$ to avoid particle buildup in the liquid film.

The video is processed using the MATLAB toolbox PIVlab 3.01. We use a multi-pass linear window deformation technique: the first pass used a 32 × 32-pixel interrogation window, the second pass 16 × 16-pixel interrogation window and sub-pixel displacement is estimated by two-dimensional Gaussian regression. Due to the sparse seeding, we adopt the algorithm of ensemble correlation to process 500 consecutive image pairs for better signal-to-noise ratio. The values of valid detection probability for all PIV results are above 0.95.

**4.4 Theoretical dominating wavelength of Plateau-Rayleigh Instability (PRI)**

The dominating wavelength for wrapping film and ridge film relies on the radius of curvature $\xi$ for fiber belly. $\xi$ is written as follows, by approximating the cross section of the fiber as an ellipse

$$\xi = \frac{a^2}{b} \tag{12}$$

For wrapping films, the equation of dominating wavelength is adopted using $\xi$ as the equivalent fiber radius, written as

$$\frac{\lambda}{\xi} = 2\pi\sqrt{2} \tag{13}$$

By substituting $\xi$ with $a$ and aspect ratio $\varepsilon$,

$$\Lambda = \frac{\lambda}{a} = \frac{2\pi\sqrt{2}}{\varepsilon} \tag{14}$$

For ridge films, the following dispersion relation derived by Diez et al.[44] is used for the growth of PRI on liquid ridge.

$$\omega = q^2 - q^4 \tag{15}$$

where $\omega$ and $q$ are the dimensionless growth rate and wave number of the perturbation. Eq. (15) gives the dominating wave number $q_d{}^2 = 1/2$ corresponding to the maximum growth rate where $d\omega/dq = 0$, and the critical wave number $q_c{}^2 = 1$ corresponding to the marginal growth rate where $\omega = 0$. Therefore, a relationship between the dominating and critical wave numbers is obtained as follows.

$$q_d = \frac{q_c}{\sqrt{2}} \tag{16}$$

As a result, the dimensional wavelength must obey the following relationship.

$$\lambda_d = \sqrt{2}\lambda_c \tag{17}$$

The critical wavelength is obtained from Langbein's stability criterion of liquid column on a solid plane with moving contact lines[45], which is written as

$$\frac{\lambda_c}{\xi_r} = 2\pi\sqrt{\frac{\pi}{\alpha}} \tag{18}$$

where $\xi_r$ is the radius of curvature of the column, and $\alpha$ is the angle between the plan and liquid-vapor interface. By considering the thin ridge film as a thin shell of liquid column, we can obtain $\xi_r \approx \xi$ and $\alpha \approx \arcsin(\varepsilon)$. Therefore, we can rewrite Eq. (18) as

$$\frac{\lambda_c}{a} = \frac{2\pi}{\varepsilon}\sqrt{\frac{\pi}{\arcsin\varepsilon}} \tag{19}$$

Using Eq. (17), we obtain the dominating wavelength as follows

$$\Lambda = \frac{\lambda_d}{a} = \frac{2\pi}{\varepsilon}\sqrt{\frac{2\pi}{\arcsin\varepsilon}} \tag{20}$$

### 4.5 Hydrodynamics of opposing flows in liquid film

For thin-film flow on ribbon-like fibers, the longitudinal component of flow velocity $u$ can be calculated using the classical Lubrication approximation by averaging the velocity along film thickness direction, as well as neglecting the inertial contribution, written as follows[26]

$$u = -\frac{h^2}{3\mu}\frac{dp}{dx} \tag{21}$$

where $h$ is the film thickness, $\mu$ is the liquid viscosity, $p$ is the pressure, and $x$ is the fiber longitudinal axis. The pressure gradient $dp/dx$ consists of (a) the capillary pressure difference between the drop and the film and (b) the capillary pressure gradient along the film due to varying film thickness. Therefore, the pressure gradient can be calculated as

$$\frac{dp}{dx} = \frac{\Delta p}{x} - \frac{\Delta p}{\lambda - x} + \gamma\frac{d\kappa}{dx} \tag{22}$$

where $\Delta p$ is the film-drop pressure difference, $\lambda$ is the PRI wavelength, and $\kappa$ is the curvature of the liquid film.

For the forward flow, the last term in Eq. (22) vanishes because the major contribution is the film-drop pressure difference. In addition, the capillary pressure is $2\gamma/r$ for PRI drop and $\gamma(2h+2b)/a^2$ for film, because the liquid-vapor interface of the film is a circular arc sitting on an elliptical fiber. Therefore, the film-drop pressure difference is obtained as follows

$$\Delta p = 2\gamma \left( \frac{h+b}{a^2} - \frac{1}{r} \right) \tag{23}$$

where $b$ and $a$ are the short (minor) and long (major) radius of the ellipse for the cross section of the fiber, and $r$ is the radius of PRI drops. Therefore, the velocity of the forward flow is calculated from

$$u_f = -\frac{2\gamma h^2}{3\mu} \left( \frac{h+b}{a^2} - \frac{1}{r} \right) \left( \frac{1}{x} - \frac{1}{\lambda - x} \right) \tag{24}$$

To solve Eq. (24), the film thickness $h$ is calculated by applying the mass conservation, where the liquid deposited on fiber equals the liquid transported to PRI drops. Therefore, the film thickness $h$ from the edge of a PRI drop ($x = r$) to the midpoint between two PRI drops ($x = \lambda/2$) follows

$$\frac{j\lambda}{2} = h\left(u_f|_{x=r} - u_f|_{x=\lambda/2}\right) \tag{25}$$

where $j$ is the flux of the droplet in the crossflow airstream that is deposited on the fiber. Substituting $u_f$ with Eq.(24) gives

$$\frac{j\mu}{\gamma} = \frac{4}{3}\frac{h^3}{\lambda} \left( \frac{h+b}{a^2} - \frac{1}{r} \right) \left( \frac{1}{r} - \frac{1}{\lambda - r} \right) \tag{26}$$

Further simplification of Eq.(26) is possible because $\lambda \gg r$.

$$\frac{j\mu}{\gamma} = \frac{4}{3}\frac{h^3}{\lambda r} \left( \frac{2h+2b}{a^2} - \frac{1}{r} \right) \tag{27}$$

The characteristic length, $a$, and the characteristic velocity, $\gamma/\mu$, are used to cast the following non-dimensional parameters:

$$H = \frac{h}{a}; X = \frac{x}{a}; U = \left|\frac{u\mu}{\gamma}\right|; \Lambda = \frac{\lambda}{a}; \varepsilon = \frac{b}{a}; R = \frac{r}{a}; J = \frac{j\mu}{\gamma} \tag{28}$$

Then, Eqs. (24) and (27) are nondimensionalized as follows.

$$U_f = \frac{2H^2}{3}\left(H + \varepsilon - \frac{1}{R}\right)\left(\frac{1}{X} - \frac{1}{\Lambda - X}\right) \tag{29}$$

$$J = \frac{4H^3}{3\Lambda R}\left(H + \varepsilon - \frac{1}{R}\right) \tag{30}$$

Here, the film thickness $H$ is the actual thickness on fiber belly.

For the backward flow, the term in Eq. (22) containing the film-drop pressure difference, $\Delta p$, vanishes due to the squeezing film. Meanwhile, the gradient term exists to generate the reverse suction. Using the dimensionless film curvature $2H+2\varepsilon$, the backward velocity can be calculated as follows

$$U_b = -\frac{2H_b^2}{3}\frac{dH_b}{dX} \tag{31}$$

Here, $H_b$ denotes the film thickness for the backward flow. For ribbon-like fibers with small $\varepsilon$, $H_b \approx H/2$ gives a good estimation (Supplementary Note 1). In addition, the gradient term $dH/dX$ due to viscous entrainment can be scaled as $dH/dX \sim H/X$, which represents an extreme case where all liquid on the fiber shoulder is entrained into the forward flow. Therefore, the backward velocity can be simplified as follows

$$U_b = \frac{H^3}{12X} \tag{32}$$

## Acknowledgment

The authors would like to thank Dr. Yifu Li and Dr. Qun Chen for proofing this manuscript and providing valuable opinions. The authors would like to acknowledge the financial support from GCI Ventures Capital, Inc., Toronto, Canada.

## Conflict of Interest

The authors declare no conflict of interest.

### References

1 Kwon, S. *et al.* Association of social distancing and face mask use with risk of COVID-19. *Nat. Commun.* **12**, 3737, doi:10.1038/s41467-021-24115-7 (2021).
2 Morawska, L., Buonanno, G., Mikszewski, A. & Stabile, L. The physics of respiratory particle generation, fate in the air, and inhalation. *Nat. Rev. Phys.* **4**, 723-734, doi:10.1038/s42254-022-00506-7 (2022).
3 Schemenauer, R. S. & Joe, P. I. The collection efficiency of a massive fog collector. *Atmos. Res.* **24**, 53-69, doi:https://doi.org/10.1016/0169-8095(89)90036-7 (1989).

4. Jacob, D. J., Wang, R. F. & Flagan, R. C. Fogwater collector design and characterization. *Environ. Sci. Technol.* **18**, 827-833, doi:10.1021/es00129a005 (1984).
5. Kern, V. R. & Carlson, A. Twisted fibers enable drop flow control and enhance fog capture. *PNAS* **121**, e2402252121, doi:10.1073/pnas.2402252121 (2024).
6. Yang, Y. *et al.* Purification technology of oil mist in industrial buildings: A review. *Build. Environ.* **235**, doi:10.1016/j.buildenv.2023.110229 (2023).
7. Wang, Y., Murga, A., Long, Z., Yoo, S.-J. & Ito, K. Experimental study of oil mist characteristics generated from minimum quantity lubrication and flood cooling. *Energy Built Environ.* **2**, 45-55, doi:10.1016/j.enbenv.2020.05.005 (2021).
8. Dbouk, T. & Drikakis, D. On respiratory droplets and face masks. *Phys. Fluids* **32**, 063303, doi:10.1063/5.0015044 (2020).
9. Mullins, B. J., Mead-Hunter, R., Pitta, R. N., Kasper, G. & Heikamp, W. Comparative performance of philic and phobic oil-mist filters. *AIChE J.* **60**, 2976-2984, doi:10.1002/aic.14479 (2014).
10. Shi, W., Anderson, M. J., Tulkoff, J. B., Kennedy, B. S. & Boreyko, J. B. Fog Harvesting with Harps. *ACS Appl. Mater. Interfaces* **10**, 11979-11986, doi:10.1021/acsami.7b17488 (2018).
11. Park, J. *et al.* Clogged water bridges for fog harvesting. *Soft Matter* **17**, 136-144, doi:10.1039/d0sm01133a (2021).
12. Mead-Hunter, R., King, A. J. & Mullins, B. J. Plateau Rayleigh instability simulation. *Langmuir* **28**, 6731-6735, doi:10.1021/la300622h (2012).
13. Haefner, S. *et al.* Influence of slip on the Plateau-Rayleigh instability on a fibre. *Nat. Commun.* **6**, 7409, doi:10.1038/ncomms8409 (2015).
14. Gao, N. *et al.* How drops start sliding over solid surfaces. *Nat. Phys.* **14**, 191-196, doi:10.1038/nphys4305 (2017).
15. Wilson, J. L. *et al.* Aerodynamic interactions of drops on parallel fibres. *Nat. Phys.* **19**, 1667-1672, doi:10.1038/s41567-023-02159-4 (2023).
16. Bintein, P.-B., Bense, H., Clanet, C. & Quéré, D. Self-propelling droplets on fibres subject to a crosswind. *Nat. Phys.* **15**, 1027-1032, doi:10.1038/s41567-019-0599-0 (2019).
17. Zheng, Y. *et al.* Directional water collection on wetted spider silk. *Nature* **463**, 640-643, doi:10.1038/nature08729 (2010).
18. Chan, T. S., Yang, F. & Carlson, A. Directional spreading of a viscous droplet on a conical fibre. *J. Fluid Mech.* **894**, doi:10.1017/jfm.2020.240 (2020).
19. Chan, T. S., Pedersen, C., Koplik, J. & Carlson, A. Film deposition and dynamics of a self-propelled wetting droplet on a conical fibre. *J. Fluid Mech.* **907**, doi:10.1017/jfm.2020.834 (2020).
20. Quéré, D. Fluid Coating on a Fiber. *Annu. Rev. Fluid Mech.* **31**, 347-384, doi:https://doi.org/10.1146/annurev.fluid.31.1.347 (1999).
21. de Gennes, P. G. Wetting: statics and dynamics. *Rev. Mod. Phys.* **57**, 827-863, doi:10.1103/RevModPhys.57.827 (1985).
22. Lee, C. L., Chan, T. S., Carlson, A. & Dalnoki-Veress, K. Multiple droplets on a conical fiber: formation, motion, and droplet mergers. *Soft Matter* **18**, 1364-1370, doi:10.1039/d1sm01462e (2022).
23. Mullins, B. J. & Kasper, G. Comment on: "Clogging of fibrous filters by liquid aerosol particles: Experimental and phenomenological modelling study" by Frising et al. *Chem. Eng. Sci.* **61**, 6223-6227, doi:10.1016/j.ces.2006.05.027 (2006).
24. Bitten, J. F. Coalescence of water droplets on single fibers. *J. Colloid Interface Sci.* **33**, 265-271, doi:https://doi.org/10.1016/0021-9797(70)90028-7 (1970).
25. Labbé, R. & Duprat, C. Capturing aerosol droplets with fibers. *Soft Matter* **15**, 6946-6951, doi:10.1039/c9sm01205b (2019).

26  de Gennes, P.-G., Brochard-Wyart, F. & Quéré, D. in *Capillarity and Wetting Phenomena: Drops, Bubbles, Pearls, Waves* (eds Pierre-Gilles de Gennes, Françoise Brochard-Wyart, & David Quéré) 107-138 (Springer New York, 2004).
27  Zhang, C., Beard, C. E., Adler, P. H. & Kornev, K. G. Effect of curvature on wetting and dewetting of proboscises of butterflies and moths. *R. Soc. Open Sci.* **5**, 171241, doi:10.1098/rsos.171241 (2018).
28  Mulji, N. & Chandra, S. Rupture and dewetting of water films on solid surfaces. *J. Colloid Interface Sci.* **352**, 194-201, doi:10.1016/j.jcis.2010.08.020 (2010).
29  Peschka, D. *et al.* Signatures of slip in dewetting polymer films. *PNAS* **116**, 9275-9284, doi:10.1073/pnas.1820487116 (2019).
30  Davis, S. H. Moving contact lines and rivulet instabilities. Part 1. The static rivulet. *J. Fluid Mech.* **98**, 225-242, doi:10.1017/S0022112080000110 (1980).
31  Quéré, D., di Meglio, J.-M. & Brochard-Wyart, F. Spreading of Liquids on Highly Curved Surfaces. *Science* **249**, 1256-1260, doi:10.1126/science.249.4974.1256 (1990).
32  Drazin, P. G. & Howard, L. N. in *Advances in Applied Mechanics* Vol. 9 (eds G. G. Chernyi *et al.*) 1-89 (Elsevier, 1966).
33  Ghasemi, A., Tuna, B. A. & Li, X. Shear/rotation competition during the roll-up of acoustically excited shear layers. *J. Fluid Mech.* **844**, 831-854, doi:10.1017/jfm.2018.214 (2018).
34  Sreenivasan, K. R. Turbulent mixing: A perspective. *PNAS* **116**, 18175-18183, doi:10.1073/pnas.1800463115 (2019).
35  Sudarsan, A. P. & Ugaz, V. M. Multivortex micromixing. *PNAS* **103**, 7228-7233, doi:10.1073/pnas.0507976103 (2006).
36  Bai, H. *et al.* Direction controlled driving of tiny water drops on bioinspired artificial spider silks. *Adv. Mater.* **22**, 5521-5525, doi:10.1002/adma.201003169 (2010).
37  Ju, J. *et al.* A multi-structural and multi-functional integrated fog collection system in cactus. *Nat. Commun.* **3**, 1247, doi:10.1038/ncomms2253 (2012).
38  Ju, J. *et al.* Cactus Stem Inspired Cone‐Arrayed Surfaces for Efficient Fog Collection. *Adv. Funct. Mater.* **24**, 6933-6938, doi:10.1002/adfm.201402229 (2014).
39  Chen, H. *et al.* Ultrafast water harvesting and transport in hierarchical microchannels. *Nat. Mater.* **17**, 935-942, doi:10.1038/s41563-018-0171-9 (2018).
40  Pan, Z. *et al.* The upside-down water collection system of Syntrichia caninervis. *Nat. Plants* **2**, 16076, doi:10.1038/nplants.2016.76 (2016).
41  Li, E. Q. & Thoroddsen, S. T. The fastest drop climbing on a wet conical fibre. *Phys. Fluids* **25**, doi:10.1063/1.4805068 (2013).
42  de Bruyn, J. R. Crossover between surface tension and gravity-driven instabilities of a thin fluid layer on a horizontal cylinder. *Phys. Fluids* **9**, 1599-1605, doi:10.1063/1.869280 (1997).
43  Huang, Y., Zhang, Y., Li, Y. & Tan, Z. Waterdrop-assisted efficient fog collection on micro-fiber grids. *Chem. Eng. J.* **481**, doi:10.1016/j.cej.2023.148423 (2024).
44  Diez, J. A., González, A. G. & Kondic, L. On the breakup of fluid rivulets. *Phys. Fluids* **21**, doi:10.1063/1.3211248 (2009).
45  Langbein, D. The shape and stability of liquid menisci at solid edges. *J. Fluid Mech.* **213**, doi:10.1017/s0022112090002312 (2006).

**Supplementary Information**

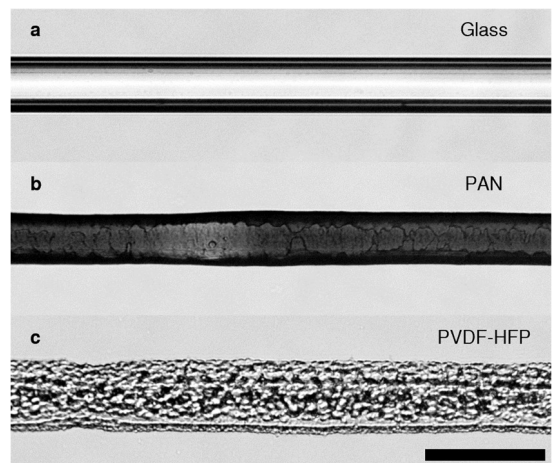

**Supplementary Fig. 8:** Microscopic image of the surface of ribbon-like fibers made of glass (a), PAN (b), and PVDF-HFP (c). Scale bar: 100 μm

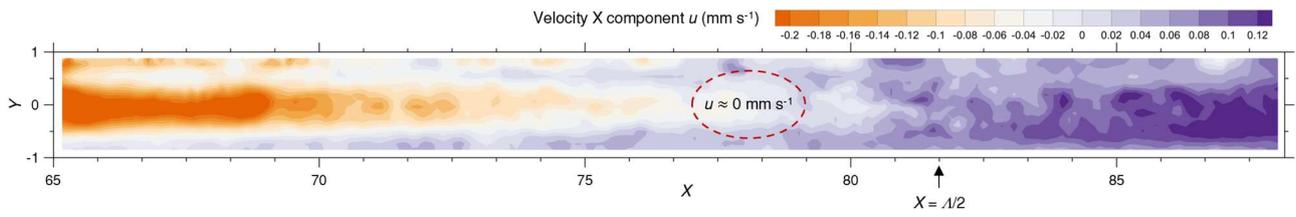

**Supplementary Fig. 9: Contour plot of PIV results for RG2, showing liquid velocity component u near X = Λ/2.** Red dashed circle indicates the position where u ≈ 0 mm s$^{-1}$.

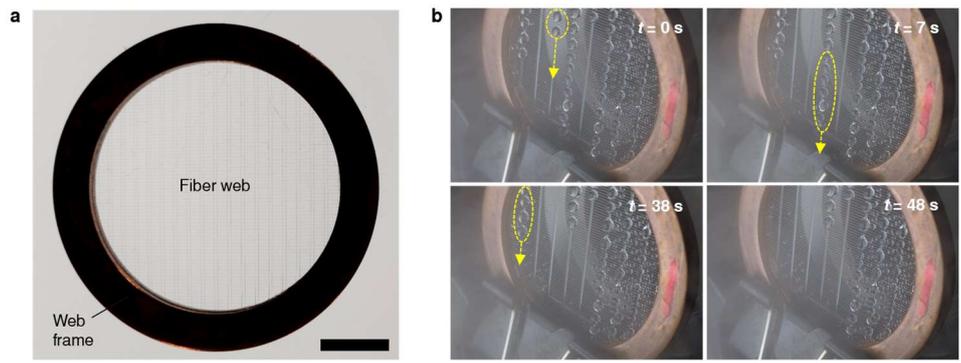

**Supplementary Fig. 10: ROW structure for continuous transport. a**, Digital image of a fiber web fixed by two copper gaskets as web frames. **b**, Digital image of the formation of isolated liquid bridges for ROW structure. Scale bar in **a**: 1 cm.

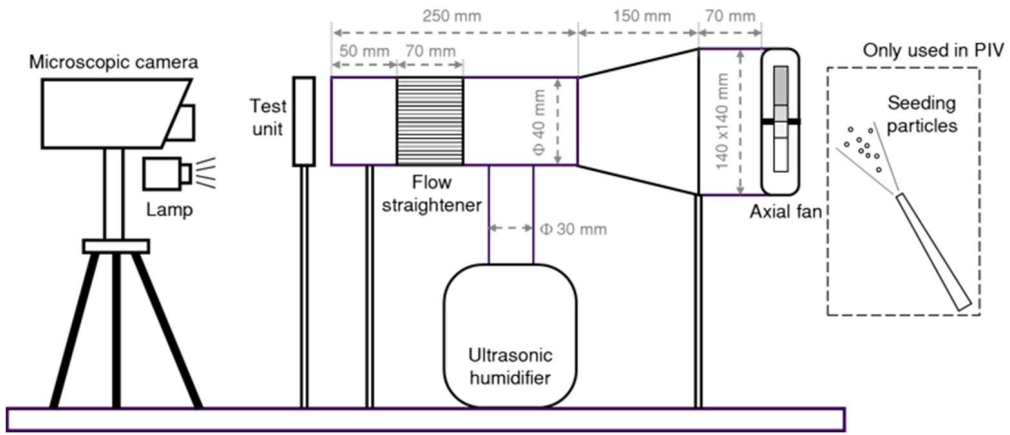

**Supplementary Fig. 11: Schematic of configurations and dimensions of the wind tunnel.**

**Supplementary Notes 1: Approximation of film thickness $H_b$**

**Supplementary Fig.** *12*a shows that the water film on a ribbon-like fiber with a major radius of $a$ and minor radius of $b$ has a cross-section shape of a circular area sitting on an ellipse. The bulging surface of the film leads to a larger belly film thickness of $h$, where the forward flow takes place, and a thinner film thickness of $h_b$, where the backward flow occurs. As discussed in the main text, the distance between $h$ and $h_b$ is about $0.7a$. However, the combination of circular and ellipse shape leads to a rather complex mathematical relationship between $h$ and $h_b$, which is difficult to interpret physically.

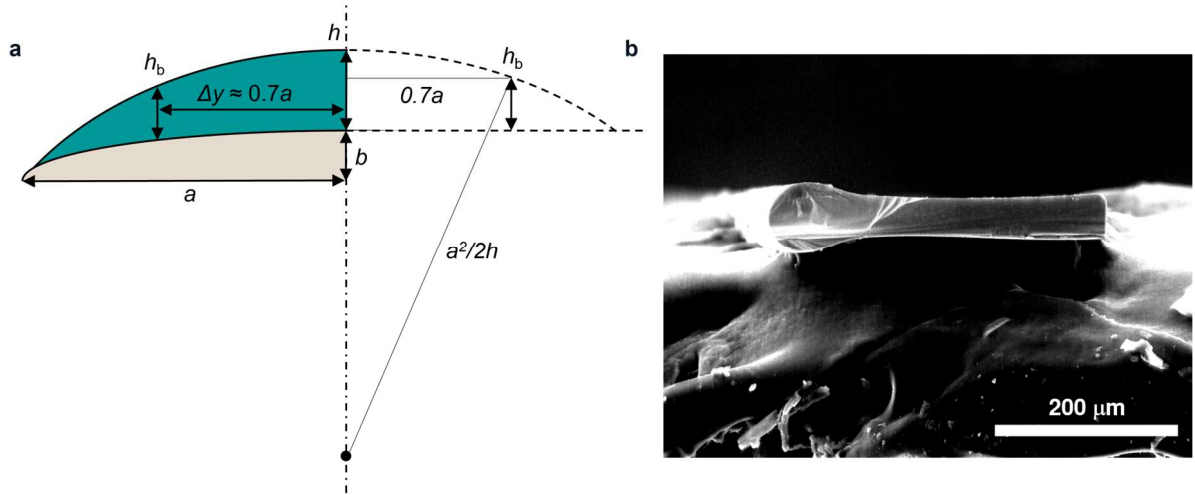

**Supplementary Fig. 12: Approximation of liquid film thickness for backward flow.** a, Schematic of approximated water film geometry. b, Scanning electron microscope image of the cross-section of a ribbon-like glass fiber with $\varepsilon = 0.2$.

Here, the geometry of the water film is approximated for better physical interpretation. Considering that $b \ll a$ for the tested ribbon-like fibers RG1 and RG2 with $\varepsilon < 0.2$, the bulging surface of the ellipse is quite flat and can be approximated as a planar surface (dash line). This approximation is practically reasonable because the ribbon-like fiber with a small $\varepsilon$ used in this study usually adopts a flatter cross section (**Supplementary Fig.** *12*b) compared to a regular ellipse. The planar surface directly simplifies the film geometry as a circular cap sitting on a plane, which leads to the following relationship between $h_b$ and $h$.

$$\frac{h_b}{h} = 1 - \left( \frac{a^2}{2h^2} - \frac{a}{h}\sqrt{\frac{a^2}{4h^2} - 0.49} \right) \tag{33}$$

The expression has a non-dimensional form as follows.

$$\frac{H_b}{H} = 1 - \left( \frac{1}{2H^2} - \frac{1}{H}\sqrt{\frac{1}{4H^2} - 0.49} \right) \tag{34}$$

Eq. (34) indicates $H_b/H$ is in the range of 0.49-0.51 for $0 < H < 0.3$, which covers the range of $H$ in this study. Therefore, the approximation provides a good estimation of $H_b$ as follows.

$$H_b = 0.5H \tag{35}$$